\documentclass[12pt]{article}
\usepackage{graphicx}

\def\pbnr{}
\def\speaker{Thomas Bird}
\def\onbehalfof{The LHCb Collaboration}
\def\title{The search for $D^0\to{}e^{\pm}\mu^{\mp}$}
\def\affiliation{School of Physics and Astronomy\\
The University of Manchester, Manchester, UK}
\def\support{}
\newcommand\pubnumber{\pbnr}
\newcommand\pubdate{\today}
\def\Title#1{\begin{center} {\Large #1 } \end{center}}
\def\Author#1{\begin{center}{ \sc #1} \end{center}}

\newcommand{\OnBehalf}[1]{\sbox0{#1}\ifdim\wd0=0pt
        {}% if #1 is empty
	\else
	{\\on behalf of #1}% if #1 is not empty
	\fi}
\newcommand{\SupportedBy}[1]{\sbox0{#1}\ifdim\wd0=0pt
        {}% if #1 is empty
	\else
	{\footnote{#1}}% if #1 is not empty
	\fi}
\def\Address#1{\begin{center}{ \it #1} \end{center}}

\newcommand\pubblock{\includegraphics[width=5cm]{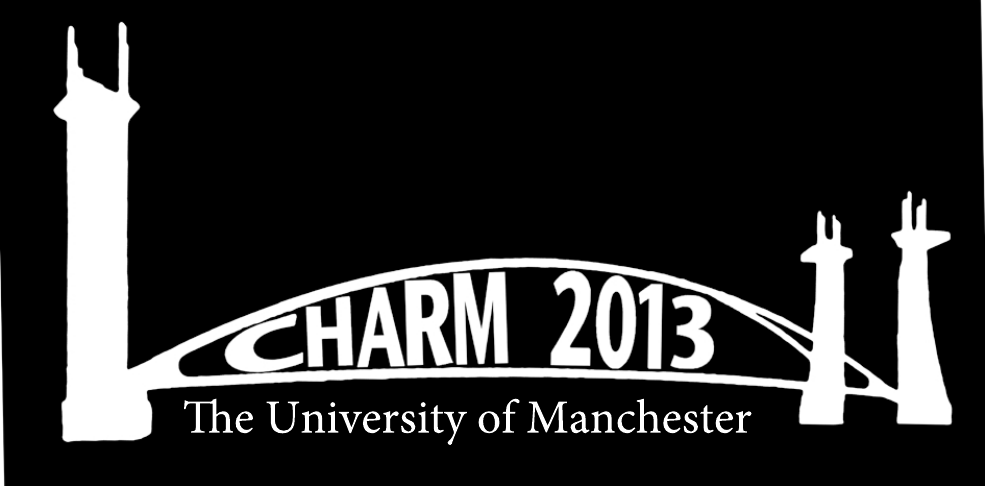}\hfill{\begin{tabular}{l} \pubnumber\\
         \pubdate  \end{tabular}}}
\newenvironment{Abstract}{\begin{quotation}  }{\end{quotation}}
\newenvironment{Presented}{\begin{quotation} \begin{center} 
             PRESENTED AT\end{center}\bigskip 
      \begin{center}\begin{large}}{\end{large}\end{center} \end{quotation}}

\def\venue{The 6$^{th}$ International Workshop on Charm Physics\\
(CHARM 2013)\\
Manchester, UK,  31 August -- 4 September, 2013}
\def\beq{\begin{equation}}
\def\eeq#1{\label{#1}\end{equation}}
\def\eeqn{\end{equation}}

\def\beqa{\begin{eqnarray}}
\def\eeqa#1{\label{#1}\end{eqnarray}}
\def\eeqan{\end{eqnarray}}

\let\bar=\overbar

\def\Dslash{\not{\hbox{\kern-4pt $D$}}}
\def\dslash{\not{\hbox{\kern-2pt $\del$}}}

\def\msb{{\bar{\ssstyle M \kern -1pt S}}}

\usepackage{ifthen} 
\newboolean{uprightparticles}
\setboolean{uprightparticles}{false} %Set true for upright particle symbols
\usepackage{xspace} 
\usepackage{upgreek}

\ifthenelse{\boolean{uprightparticles}}%
{

 \def\PDelta      {\ensuremath{\Delta}\xspace}                 
 \def\PXi      {\ensuremath{\Xi}\xspace}                 
 \def\PLambda      {\ensuremath{\Lambda}\xspace}                 
 \def\PSigma      {\ensuremath{\Sigma}\xspace}                 
 \def\POmega      {\ensuremath{\Omega}\xspace}                 
 \def\PUpsilon      {\ensuremath{\Upsilon}\xspace}                 
 
 \def\PB      {\ensuremath{\mathrm{B}}\xspace}                 
                  
 \def\PD      {\ensuremath{\mathrm{D}}\xspace}

 \def\PK      {\ensuremath{\mathrm{K}}\xspace}

 \def\Pi      {\ensuremath{\mathrm{i}}\xspace}

}
{

 \mathchardef\PDelta="7101
 \mathchardef\PXi="7104
 \mathchardef\PLambda="7103
 \mathchardef\PSigma="7106
 \mathchardef\POmega="710A
 \mathchardef\PUpsilon="7107
                  
 \def\PB      {\ensuremath{B}\xspace}                 
                  
 \def\PD      {\ensuremath{D}\xspace}

 \def\PK      {\ensuremath{K}\xspace}

 \def\Pi      {\ensuremath{i}\xspace}

}

  \def\Kbar  {\kern 0.2em\overline{\kern -0.2em \PK}{}\xspace}

  \def\Dbar    {\kern 0.2em\overline{\kern -0.2em \PD}{}\xspace}

\def\Bbar    {\ensuremath{\kern 0.18em\overline{\kern -0.18em \PB}{}}\xspace}

  \def\Y#1S{\ensuremath{\PUpsilon{(#1S)}}\xspace}% no space before {...}!

\def\Lbar {\ensuremath{\kern 0.1em\overline{\kern -0.1em\PLambda}}\xspace}

\def\to                 {\ensuremath{\rightarrow}\xspace}

\def\AT#1     {\ensuremath{A_{\mathrm{T}}^{#1}}\xspace}           % 2

\def\C#1      {\ensuremath{\mathcal{C}_{#1}}\xspace}                       % 9
\def\Cp#1     {\ensuremath{\mathcal{C}_{#1}^{'}}\xspace}                    % 7
\def\Ceff#1   {\ensuremath{\mathcal{C}_{#1}^{\mathrm{(eff)}}}\xspace}        % 9  
\def\Cpeff#1  {\ensuremath{\mathcal{C}_{#1}^{'\mathrm{(eff)}}}\xspace}       % 7
\def\Ope#1    {\ensuremath{\mathcal{O}_{#1}}\xspace}                       % 2
\def\Opep#1   {\ensuremath{\mathcal{O}_{#1}^{'}}\xspace}                    % 7

\newcommand{\tev}{\ifthenelse{\boolean{inbibliography}}{\ensuremath{~T\kern -0.05em eV}\xspace}{\ensuremath{\mathrm{\,Te\kern -0.1em V}}\xspace}}
\newcommand{\gev}{\ensuremath{\mathrm{\,Ge\kern -0.1em V}}\xspace}
\newcommand{\mev}{\ensuremath{\mathrm{\,Me\kern -0.1em V}}\xspace}
\newcommand{\kev}{\ensuremath{\mathrm{\,ke\kern -0.1em V}}\xspace}
\newcommand{\ev}{\ensuremath{\mathrm{\,e\kern -0.1em V}}\xspace}
\newcommand{\gevc}{\ensuremath{{\mathrm{\,Ge\kern -0.1em V\!/}c}}\xspace}
\newcommand{\mevc}{\ensuremath{{\mathrm{\,Me\kern -0.1em V\!/}c}}\xspace}
\newcommand{\gevcc}{\ensuremath{{\mathrm{\,Ge\kern -0.1em V\!/}c^2}}\xspace}
\newcommand{\gevgevcccc}{\ensuremath{{\mathrm{\,Ge\kern -0.1em V^2\!/}c^4}}\xspace}
\newcommand{\mevcc}{\ensuremath{{\mathrm{\,Me\kern -0.1em V\!/}c^2}}\xspace}

\def\gsim{{~\raise.15em\hbox{$>$}\kern-.85em
          \lower.35em\hbox{$\sim$}~}\xspace}
\def\lsim{{~\raise.15em\hbox{$<$}\kern-.85em
          \lower.35em\hbox{$\sim$}~}\xspace}

\def\evtgen     {\mbox{\textsc{EvtGen}}\xspace}

\def\geant      {\mbox{\textsc{Geant4}}\xspace}

\def\photos     {\mbox{\textsc{Photos}}\xspace}

\def\pythia     {\mbox{\textsc{Pythia}}\xspace}

\def\tell1  {TELL1\xspace}
\def\ukl1   {UKL1\xspace}

\usepackage{booktabs}
\usepackage{amsmath}
\usepackage{lineno}
\usepackage{graphicx}
\usepackage{caption}
\usepackage{subcaption}
\usepackage{sidecap}

\newcommand*\patchAmsMathEnvironmentForLineno[1]{%
\expandafter\let\csname old#1\expandafter\endcsname\csname #1\endcsname
\expandafter\let\csname oldend#1\expandafter\endcsname\csname
end#1\endcsname
 \renewenvironment{#1}%
   {\linenomath\csname old#1\endcsname}%
   {\csname oldend#1\endcsname\endlinenomath}%
}
\newcommand*\patchBothAmsMathEnvironmentsForLineno[1]{%
  \patchAmsMathEnvironmentForLineno{#1}%
  \patchAmsMathEnvironmentForLineno{#1*}%
}
\AtBeginDocument{%
\patchBothAmsMathEnvironmentsForLineno{equation}%
\patchBothAmsMathEnvironmentsForLineno{align}%
\patchBothAmsMathEnvironmentsForLineno{flalign}%
\patchBothAmsMathEnvironmentsForLineno{alignat}%
\patchBothAmsMathEnvironmentsForLineno{gather}%
\patchBothAmsMathEnvironmentsForLineno{multline}%
}

\usepackage[
    pdftitle={The search for D to e mu},
    pdfauthor={Thomas Bird},
    pdfkeywords={LHCb, D-mesons, LFV},
    bookmarksnumbered,
    colorlinks=true,
    allcolors=black,
]{hyperref} 
\makeatletter

\renewcommand{\SC@figure@vpos}{b}
\makeatother

\newcommand{\figuredir}{.}

\begin{document}
\begin{titlepage}
\pubblock

\vfill
\Title{\title}
\vfill
\Author{\speaker\SupportedBy{\support}\OnBehalf{\onbehalfof}}
\Address{\affiliation}
\vfill
\begin{Abstract}
%%%%%%%%%%%%%%%%%%%%%%%%%%%%%%%%%%%%%%%%%%%%%%%%%%%%%%%%%%%%%%%%%%%%%%%%%%%
% YOUR ABSTRACT GOES HERE
%%%%%%%%%%%%%%%%%%%%%%%%%%%%%%%%%%%%%%%%%%%%%%%%%%%%%%%%%%%%%%%%%%%%%%%%%%%

An overview is presented of a method to search for $D^0\to{}e^{\pm}\mu^{\mp}$ with LHCb data.
In order to reduce combinatorial backgrounds, tagged $D^0$ candidates from the decay $D^{\ast+}\to{}D^0\pi^+$ are used.
This measurement is performed with respect to $\mathcal{B}\left(D^0\to{}\pi^+\pi^-\right)$, which cancels uncertainties in the luminosity and $D^{\ast+}$ production cross-section.
It is estimated that using $3\,\mathrm{fb}^{-1}$ of LHCb data an upper limit can be attained of $\mathcal{O}\left(10^{-7}\right)$ at a $90\%$ confidence level.

\end{Abstract}
\vfill
\begin{Presented}
\venue
\end{Presented}
\vfill
\end{titlepage}
\def\thefootnote{\fnsymbol{footnote}}
\setcounter{footnote}{0}
%

% \linenumbers

%%%%%%%%%%%%%%%%%%%%%%%%%%%%%%%%%%%%%%%%%%%%%%%%%%%%%%%%%%%%%%%%%%%%%%%%%%%
%  WHAT FOLLOWS IS YOUR TEXT
%%%%%%%%%%%%%%%%%%%%%%%%%%%%%%%%%%%%%%%%%%%%%%%%%%%%%%%%%%%%%%%%%%%%%%%%%%%
\section{Overview}

The lepton flavour violating decay $D^0\to{}e^{\pm}\mu^{\mp}$ \footnote{In this paper, charge conjugate modes are always implied.} is forbidden in the Standard Model (SM) and so its observation would be a clear sign of new physics.
This decay is predicted to occur by several different SM extensions, with predicted rates varying by eight orders of magnitude.
An $R$-parity violating minimal supersymmetric SM predicts $\mathcal{B}\left(D^0\to{}e^{\pm}\mu^{\mp}\right)<1.0\times{}10^{-6}$~\cite{Burdman:2001tf}, while theories with multiple Higgs doublets predict rates $4$ orders of magnitude lower, $\mathcal{B}\left(D^0\to{}e^{\pm}\mu^{\mp}\right)\sim7\times{}10^{-10}$~\cite{Burdman:2001tf}. The lowest branching fraction predictions are given by SM extensions with extra fermions, which predict $\mathcal{B}\left(D^0\to{}e^{\pm}\mu^{\mp}\right)<1.0\times{}10^{-14}$~\cite{Burdman:2001tf}.

In 2011 and 2012 the LHCb experiment collected $3\,\textrm{fb}^{-1}$ of $pp$ collisions produced by the LHC at center-of-mass energies $\sqrt{s}=7\,\textrm{TeV}$ and $8\,\textrm{TeV}$.
The analysis procedure is presented to measure $\mathcal{B}\left(D^0\to{}e^{\pm}\mu^{\mp}\right)$ with the $3\,\textrm{fb}^{-1}$ dataset.
In order to reduce the combinatorial background, tagged $D^0$ candidates are used from the decay $D^{\ast+}\to{}D^0\pi^+$.
The $\mathcal{B}\left(D^0\to{}e^{\pm}\mu^{\mp}\right)$ measurement will be performed with respect to $\mathcal{B}\left(D^0\to{}\pi^+\pi^-\right)$, in order to cancel uncertainties in the luminosity and $D^{\ast+}$ production cross-section.

Candidate $D^0\to{}e^{\pm}\mu^{\mp}$ decays are selected in bins of Boosted Decision Tree (BDT) output; this helps to separate signal from background events.
The number of signal and background events are fitted in three BDT bins and combined with the PDG value of $\mathcal{B}\left(D^0\to{}\pi^+\pi^-\right)$, which is currently $\left(1.402\pm0.026\right)\times{}10^{-3}$~\cite{Beringer:1900zz}, to produce a limit on $\mathcal{B}\left(D^0\to{}e^{\pm}\mu^{\mp}\right)$.

\section{Selection efficiency}

Before the total number of $D^0\to{}e^{\pm}\mu^{\mp}$ or $D^0\to{}\pi^+\pi^-$ events can be measured the selection efficiency must be measured; this is performed on simulated events.
\pythia~6.4~\cite{Sjostrand:2006za} is used for the simulation, with a specific LHCb configuration~\cite{Belyaev:2010yga}.
Decays of hadronic particles are described by \evtgen~\cite{Lange:2001uf}, in which final-state radiation is generated using \photos~\cite{Golonka:2005pn}.
Particle interactions with the detector and the detector response are implemented using the \geant toolkit~\cite{Allison:2006ve, Agostinelli:2002hh} as described in Ref.~\cite{Nicol:2012np}.
Identical reconstruction and selection algorithms can be applied to both the data and simulation.%, which ensures the selection efficiency measured in simulation is as similar as possible to that in data.

The selection efficiency, $\varepsilon_{\mathrm{sel}}$, can be broken down into several parts, $$\varepsilon_{\mathrm{sel}} = {\varepsilon_{\mathrm{gen}}}~\varepsilon_{\mathrm{off|gen}}~\varepsilon_{\mathrm{trig|off}},$$ where $\varepsilon_{\mathrm{gen}}$ is the fraction of generated events that pass requirements on the muon and electron momenta during the simulation (these requirements are not applied to the $D^0\to{}\pi^+\pi^-$ sample); $\varepsilon_{\mathrm{off|gen}}$ is the fraction of generated events that pass the offline selection and $\varepsilon_{\mathrm{trig|off}}$ is the fraction of simulated events that are triggered given they passed the offline selection.
The measured values of these efficiencies are given in Table~\ref{Table:efficiency}.
\begin{table}[t]
\begin{center}
\caption{Generation, offline and trigger efficiencies measured on simulated LHCb data for the decays $D^0\to{}e^{\pm}\mu^{\mp}$ and $D^0\to{}\pi^+\pi^-$.}
\vspace{0.5em}
\begin{tabular}{lccc}
\toprule
                                 & Generation & Offline & Trigger \\
Decay                            &  efficiency, $\varepsilon_{\mathrm{gen}}$ &  efficiency, $\varepsilon_{\mathrm{off|gen}}$ &  efficiency, $\varepsilon_{\mathrm{trg|off}}$ \\ \midrule
 $D^0\to{}e^{\pm}\mu^{\mp}$   &  $14\%$       &  $1.6\%$       &   $19\%$ \\
 $D^0\to{}\pi^+\pi^-$ &  $100\%$      &  $1.8\%$       &   $20\%$ \\ \bottomrule
\end{tabular}
\label{Table:efficiency}
\end{center}
\end{table}

\begin{figure}[b]
%   \includegraphics[keepaspectratio,width=.3\textwidth]{\figuredir/overtraining_BDT_ada.pdf}
%   \caption{\tiny{}BDT output for background (blue) and signal (red) events. Events to the left are more likely background and events to the right are more likely signal. Independent test samples of signal and background events were used to check the BDT was not overtrained; the output of a signal (background) test sample can be seen as yellow (green) circles.}\label{Figure:BDT}
  \begin{subfigure}[c]{0.49\textwidth}
    \centering
    \includegraphics[keepaspectratio,width=\linewidth]{\figuredir/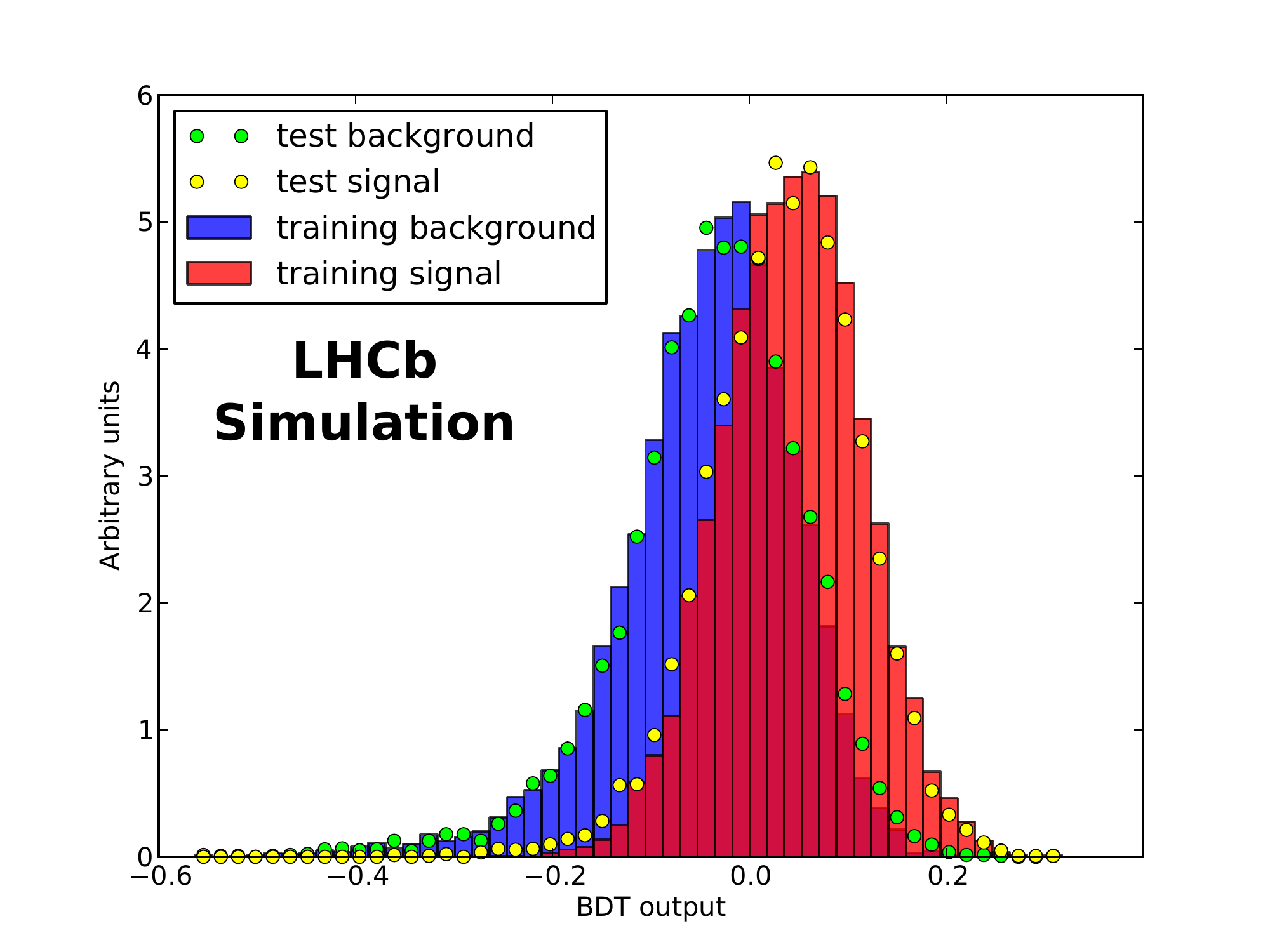}
  \end{subfigure}
  \begin{minipage}[c]{0.48\textwidth}
    \caption{BDT output for background (blue) and signal (red) events. Events to the left are more likely background and events to the right are more likely signal. Independent test samples of signal and background events were used to check the BDT was not overtrained; the output of a signal (background) test sample can be seen as yellow (green) circles.}\label{Figure:BDT}
  \end{minipage}
\end{figure}

\section{Signal-background separation}

\begin{figure}[!ht]
  \centering
  \begin{subfigure}[c]{0.49\textwidth}
    \centering
    \includegraphics[keepaspectratio,width=.9\linewidth]{\figuredir/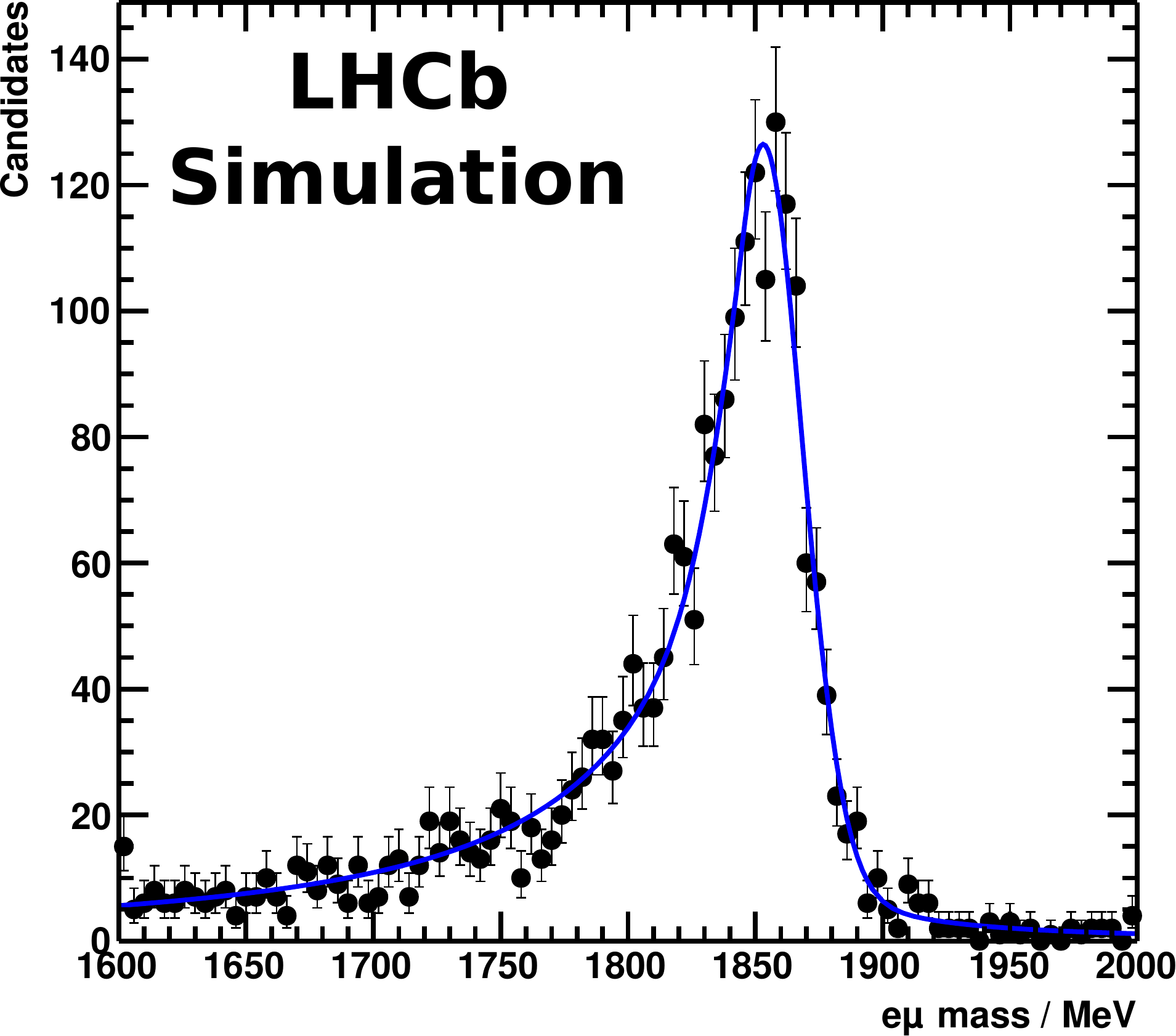}
  \end{subfigure}
  \begin{subfigure}[c]{0.49\textwidth}
    \centering
    \includegraphics[keepaspectratio,width=.9\linewidth]{\figuredir/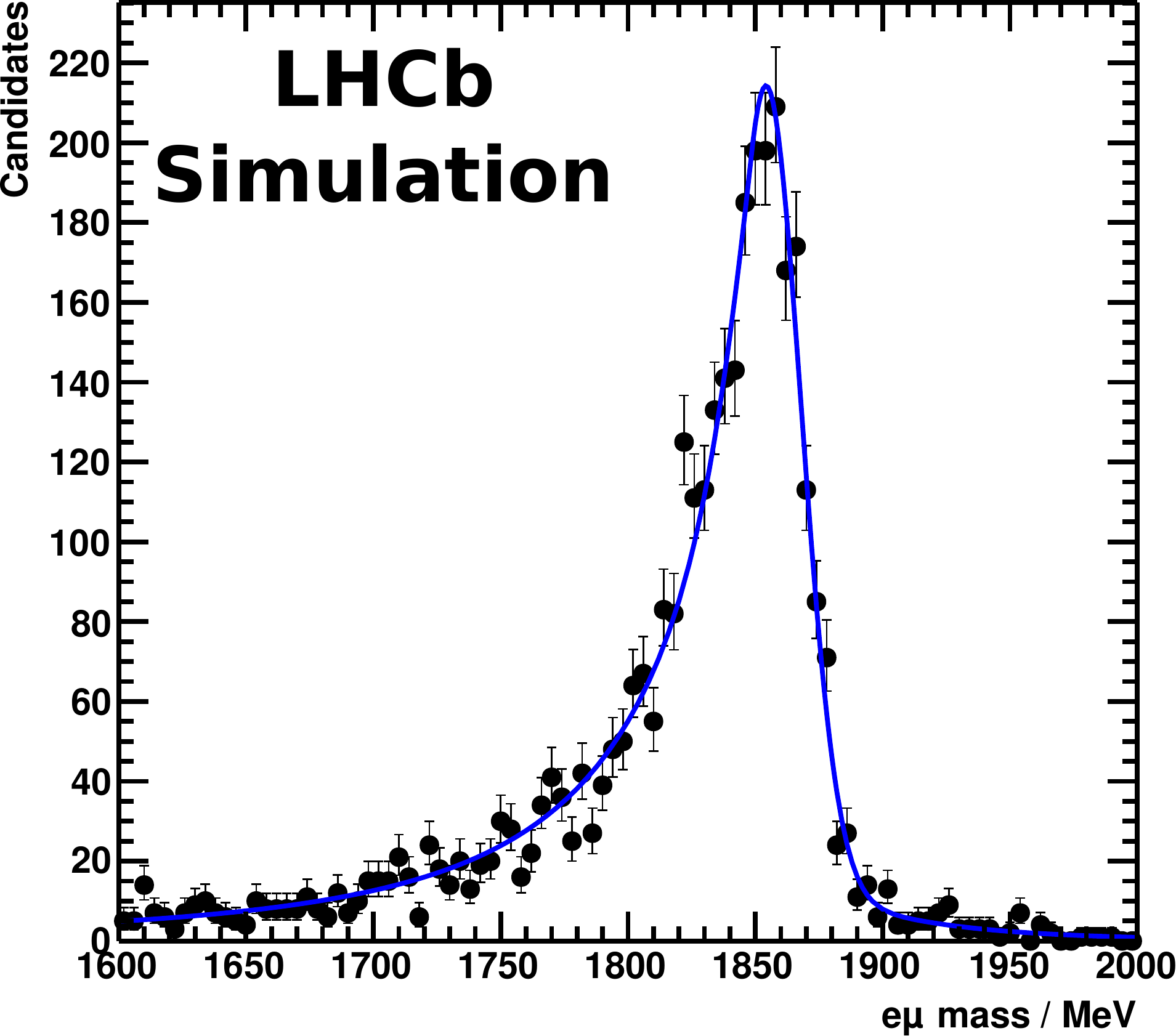}
  \end{subfigure}

  \begin{subfigure}[c]{0.49\textwidth}
    \vspace{2em}
    \centering
    \includegraphics[keepaspectratio,width=.9\linewidth]{\figuredir/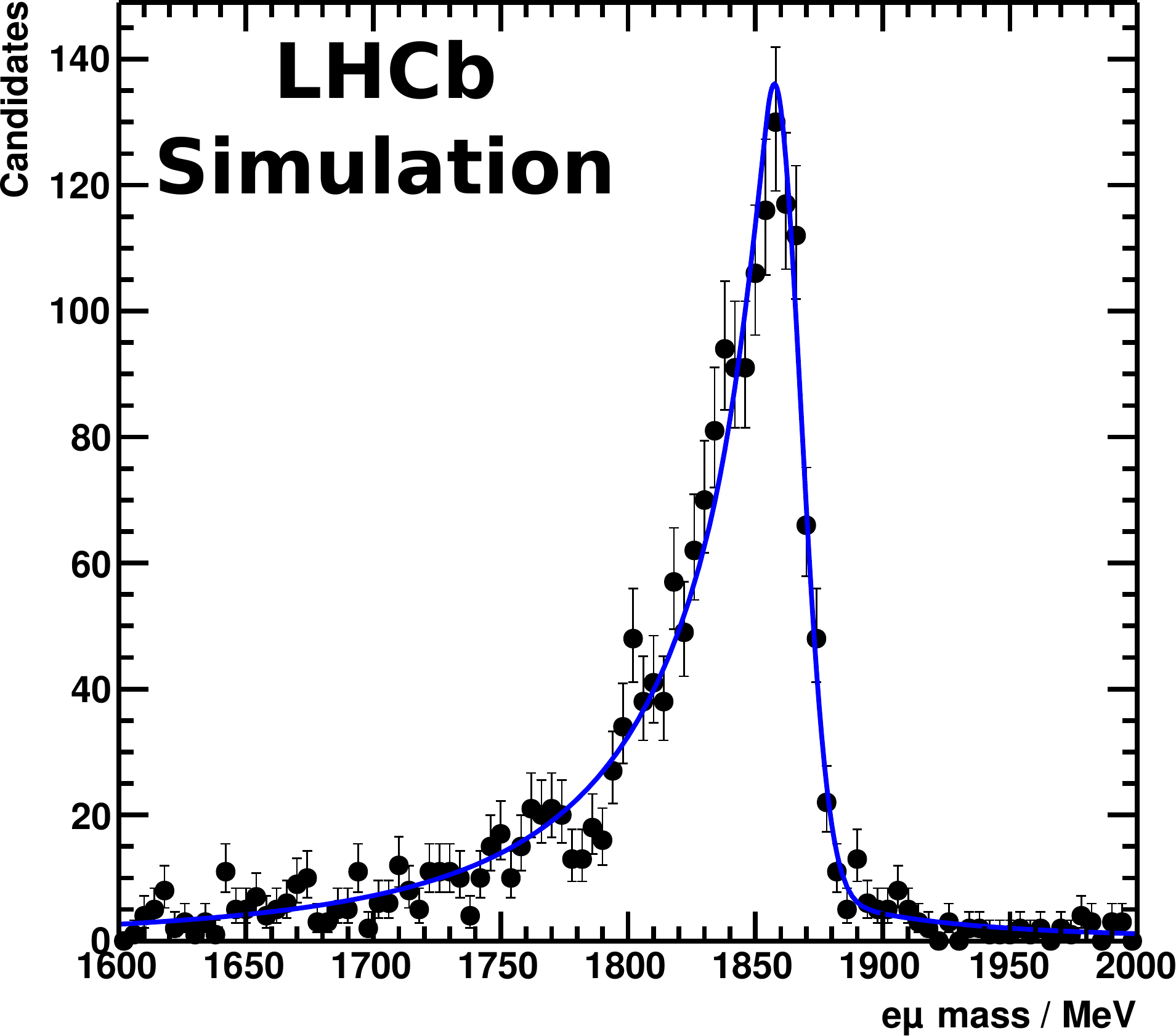}
  \end{subfigure}
  \begin{minipage}[c]{0.48\textwidth}
    \vspace{2em}
    \caption{Mass distribution of the $e\mu$-combination from simulated $D^0\to{}e^{\pm}\mu^{\mp}$ data. The blue line shows the PDF used to fit the data. The three plots are the three bins of BDT output, where the plot on the top (bottom) left corresponds to the bin where the lowest (highest) signal over background ratio is expected in data.}\label{Figure:MassFit}
  \end{minipage}
\end{figure}

To improve separation between signal and background events, a BDT is trained using the TMVA package~\cite{Hocker:2007ht}.
The BDT is trained on several kinematic and geometrical variables using data from the $e\mu$-combination mass sidebands as a background sample and the simulated signal dataset as the signal sample; its separation power can be seen in Fig.~\ref{Figure:BDT}, where the BDT output distribution is shown for both signal and background samples.
The BDT output distribution from the signal sample is divided into three bins ($<0$, $0$--$0.08$ and $>0.08$) with an approximately equal number of events in each.
In each bin a fit is performed to two variables, the mass of the $e\mu$-combination and the mass difference between the $e\mu\pi$-combination and the $e\mu$-combination.
When fitting data, the second variable helps to separate out combinatorial background events, where a real $\pi^+$ is randomly combined with a real $D^0$ to form a fake $D^{\ast+}$.
The plots in Fig.~\ref{Figure:MassFit} show the mass distribution of the $e\mu$-combination in simulated $D^0\to{}e^{\pm}\mu^{\mp}$ events for each bin of the BDT output.

\begin{figure}[!ht]
  \centering
  \begin{subfigure}[c]{0.49\textwidth}
    \centering
    \includegraphics[keepaspectratio,width=.9\linewidth]{\figuredir/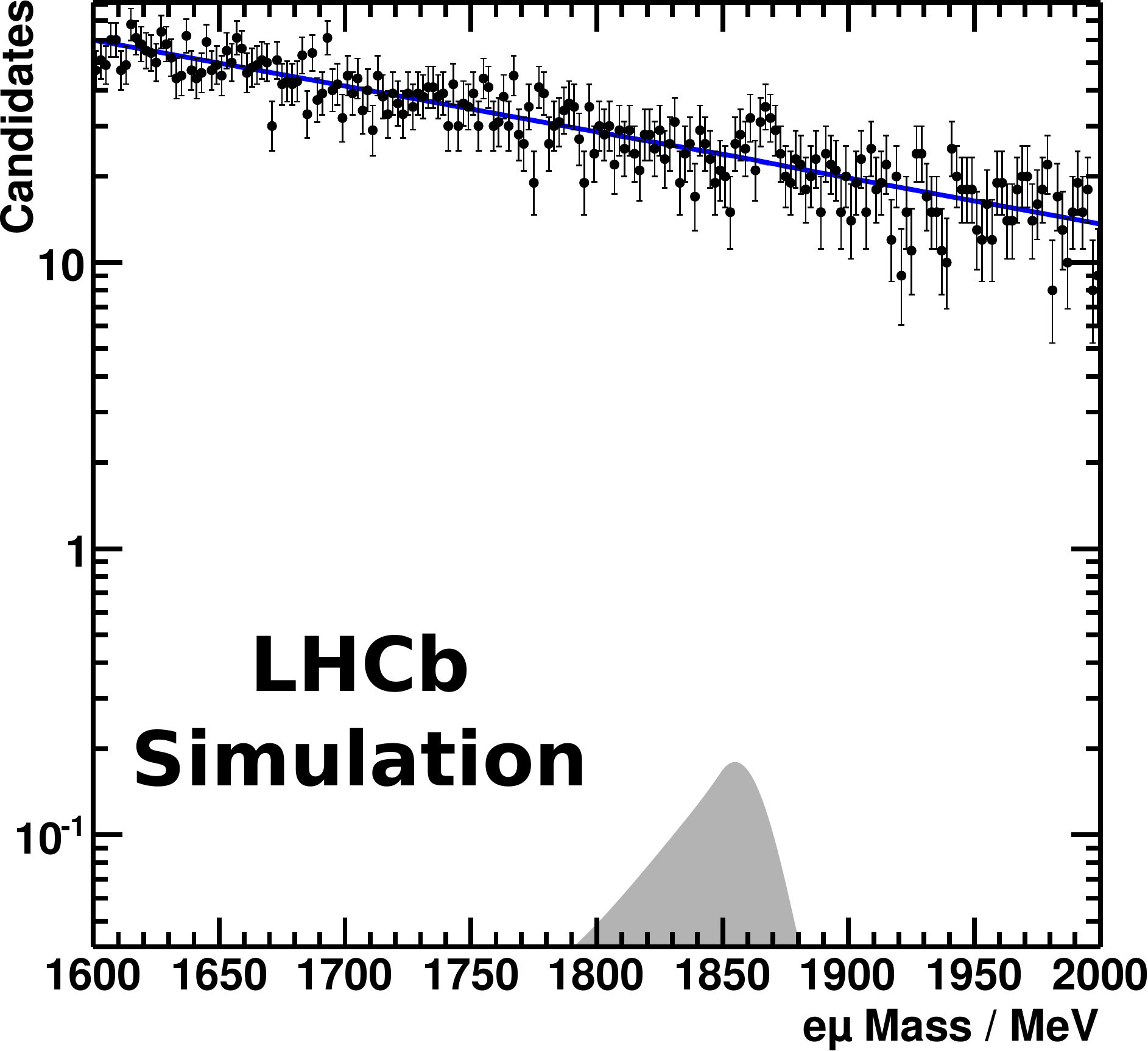}
  \end{subfigure}
  \begin{subfigure}[c]{0.49\textwidth}
    \centering
    \includegraphics[keepaspectratio,width=.9\linewidth]{\figuredir/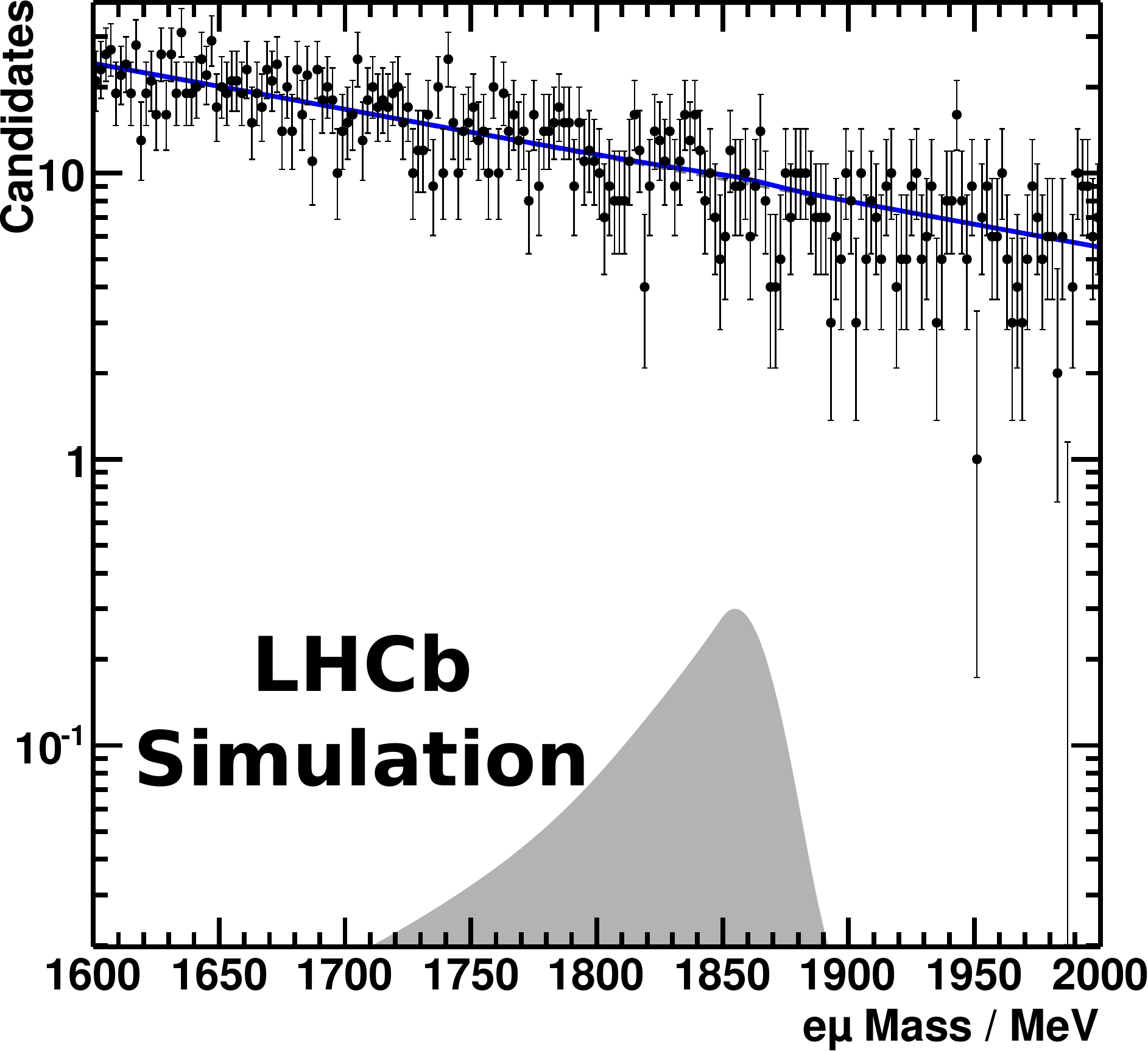}
  \end{subfigure}

  \begin{subfigure}[c]{0.49\textwidth}
    \vspace{2em}
    \centering
    \includegraphics[keepaspectratio,width=.9\linewidth]{\figuredir/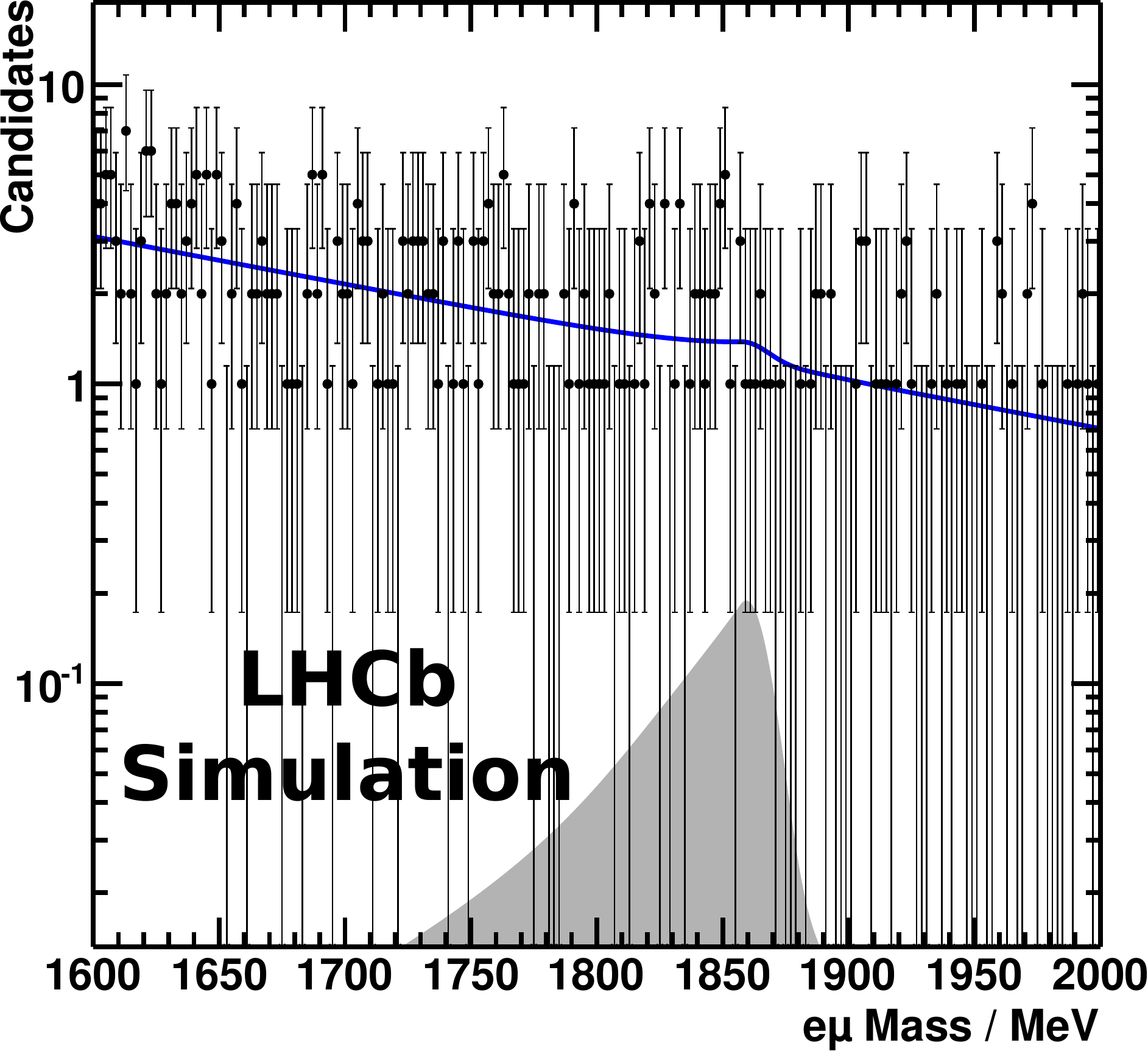}
  \end{subfigure}
  \begin{minipage}[c]{0.49\textwidth}
    \vspace{2em}
    \caption{Mass distribution of the $e\mu$-combination from simulated toy background data. The blue line shows the overall fit model used; it contains both a background and signal (solid grey area) component. The three plots are the three bins of BDT output, where the plot on the top (bottom) left corresponds to the bin where the lowest (highest) signal over background ratio is expected in data.}\label{Figure:Toy}
  \end{minipage}
\end{figure}

\section{Estimation of expected limit}

In order to estimate the possible limit that could be achieved on $\mathcal{B}\left(D^0\to{}e^{\pm}\mu^{\mp}\right)$ a toy study is performed.
In this toy study the $D^0\to{}\pi^+\pi^-$ events are simultaneously fitted along with $D^0\to{}e^{\pm}\mu^{\mp}$ events split into the three BDT bins.
The number of expected signal $D^0\to{}\pi^+\pi^-$ events after selection is approximated using the known $D^{\ast+}$ cross-section, branching ratios, integrated luminosity and measured efficiencies; it is calculated approximately $51000$ $D^0\to{}\pi^+\pi^-$ events will be selected in $3\,\textrm{fb}^{-1}$ of LHCb data.
No signal $D^0\to{}e^{\pm}\mu^{\mp}$ events are simulated in this study, but they are fitted for nevertheless to evaluate the expected sensitivity in data.
The signal $D^0\to{}e^{\pm}\mu^{\mp}$ shape is taken and fixed from the fit to the simulated data, while for the background shape various assumptions are made about the number and distribution of events.
Figure~\ref{Figure:Toy} shows the $e\mu$-combination mass distribution in the three bins of BDT output.
The toy study is able to set an upper limit on the branching fraction $\mathcal{B}\left(D^0\to{}e^{\pm}\mu^{\mp}\right)<\mathcal{O}\left(10^{-7}\right)$ at a $90\%$ confidence level.
The upper limit is calculated by taking the ratio of the signal-plus-background and the background-only profile likelihoods and assumes Wilks' theorem~\cite{Wilks:1938dza} to calculate a $p$-value and thus the confidence interval.

% \section{Future Limit}

Belle measured the most stringent limit on the branching fraction $\mathcal{B}\left(D^0\to{}e^{\pm}\mu^{\mp}\right)$, achieving an upper limit of $2.6\times{}10^{-7}$ at a $90\%$ confidence level \cite{Petric:2010yt}.
% Using the $3\,\mathrm{fb}^{-1}$ dataset, the single event sensitivity in LHCb is estimated to be $2\times{}10^{-8}$.
It is likely LHCb will set a comparable limit with the $3\,\textrm{fb}^{-1}$ dataset.

% \Acknowledgements
% I am grateful to Don Alfonso d'Alba for certain services essential to
% this investigation.

\end{document}